\documentclass[pra,onecolumn,floatfix,showpacs, showkeys]{revtex4}
\usepackage[pdftex]{graphicx}
\usepackage{bm}
\usepackage{amsmath}
\usepackage{amssymb}
\usepackage{mathrsfs}
\usepackage{array}
\usepackage{tabulary}

\newcolumntype{K}[1]{>{\centering\arraybackslash}p{#1}}

\renewcommand{\deg}{$^{\circ}$}

\newcommand{\newc}{\newcommand}
 
\newc{\be}{\begin{equation}}
\newc{\ee}{\end{equation}}
\newc{\bea}{\begin{eqnarray}}
\newc{\eea}{\end{eqnarray}}
\newc{\ie}{{\it i.e.} }
\newc{\eg}{{\it e.g.} }
\newc{\etc}{{\it etc.} }
\newc{\etal}{{\it et al.} } 
\newc{\ra}{\rightarrow}
\newc{\lra}{\leftrightarrow}
\newc{\lsim}{\buildrel{<}\over{\sim}}
\newc{\gsim}{\buildrel{>}\over{\sim}}

\begin{document}

\title{Symmetry and Piezoelectricity: Evaluation of $\alpha$-Quartz coefficients}

\author{C. Tannous}
\affiliation{Laboratoire des Sciences et Techniques de l'Information, de la Communication et 
de la Connaissance, UMR-6285 CNRS, Brest Cedex3, FRANCE
\thanks{Tel.: (33) 2.98.01.62.28,  E-mail: tannous@univ-brest.fr}}

\begin{abstract}
Piezoelectric coefficients of $\alpha$-Quartz are derived from symmetry arguments 
based on Neumann's Principle with three different methods: 
Fumi, Landau-Lifshitz and Royer-Dieulesaint. While Fumi method is tedious and Landau-Lifshitz
requires additional physical principles to evaluate the piezoelectric coefficients,
Royer-Dieulesaint is the most elegant and most efficient of the three techniques.
\end{abstract}

\maketitle

\section{Introduction and motivation}
Physics students are exposed to various types of symmetry~\cite{Gross}
and conservation laws in Graduate/Undergraduate Mechanics 
and Electromagnetism with Lorentz transformation and Gauge symmetries, 
in Graduate/Undergraduate Quantum Mechanics
during the study of Atoms and Molecules. In undergraduate
courses such as Special Relativity, Lorentz Transformation is used to unify
symmetries between Mechanics and Electromagnetism.

In Graduate High Energy Physics, the CPT theorem where 
C denotes charge conjugation  $(Q \rightarrow -Q)$, P is parity $(\bm{r} \rightarrow \bm{-r})$
and T is time reversal $(t \rightarrow -t)$  as well as Gauge symmetry 
($A_i \rightarrow A_i+\partial_i \chi)$ provide
an important insight into the role of symmetry in the building blocks
of matter and unification of fundamental forces and interaction between particles.

Graduate/undergraduate Solid State Physics provide a direct illustration of 
how Crystal Symmetry plays a fundamental role in the determination of 
physical constants and transport coefficients as well as conservation and simplification of 
physical laws. The relation between symmetry and dispersion
relations through Kramers theorem (T symmetry) is another example of the power of 
symmetry in Solid State physics. In Graduate/undergraduate Statistical Physics students are
exposed to the role of symmetry and its breaking in phase transitions 
with existence of different phases while possessing different symmetries 
are each characterized by an order parameter that controls the behaviour of 
the corresponding free energy.

The emergence on symmetry in physical systems in not obvious, however a good 
starting point to understand this particular point is through crystal symmetry paradigm
simply illustrated with ice formation by slowly cooling liquid water.

This work about the role of crystalline symmetries and their role in 
the determination of piezoelectric coefficients $d$  of $\alpha$-Quartz on the basis of
three different methods. It could be used to illustrate the role of symmetry and its implications
in an undergraduate or graduate Solid State Physics, Statistical Physics or Materials Science course.

It is organized as follows. After reviewing $\alpha$-Quartz properties and symmetries,
we tackle the evaluation of $d$ coefficients with symmetry on the basis of Fumi method. 
In section 3 we treat the same problem by Landau-Lifshitz method that contains a more 
physical approach than Fumi and finally in Section 4 we tackle it with a special method,
the Dieulesaint-Royer procedure that combines both previous approaches. The appendix contains 
detailed information about Point Groups and Symmetry operations.

\section{$\alpha$-Quartz symmetries}

Quartz is a very important material from the technological point of view since it is an essential
component of all oscillators (clocks) used in consumer electronics devices (watches, computers,
resonators, cameras, ovens...). Quartz is the second most important material after
Silicon. Its formula is silicon dioxide SiO$_2$ and its  solid state unit cell is shown in fig.1. 

$\alpha$-Quartz has a trigonal structure (rhombohedral~\cite{Landau}) belonging to
${\bm D}_3$ point symmetry group (Schoenflies classification) or 32 (Hermann-Mauguin or International
classification).

\begin{figure}[htbp]
\includegraphics[width=2.5in]{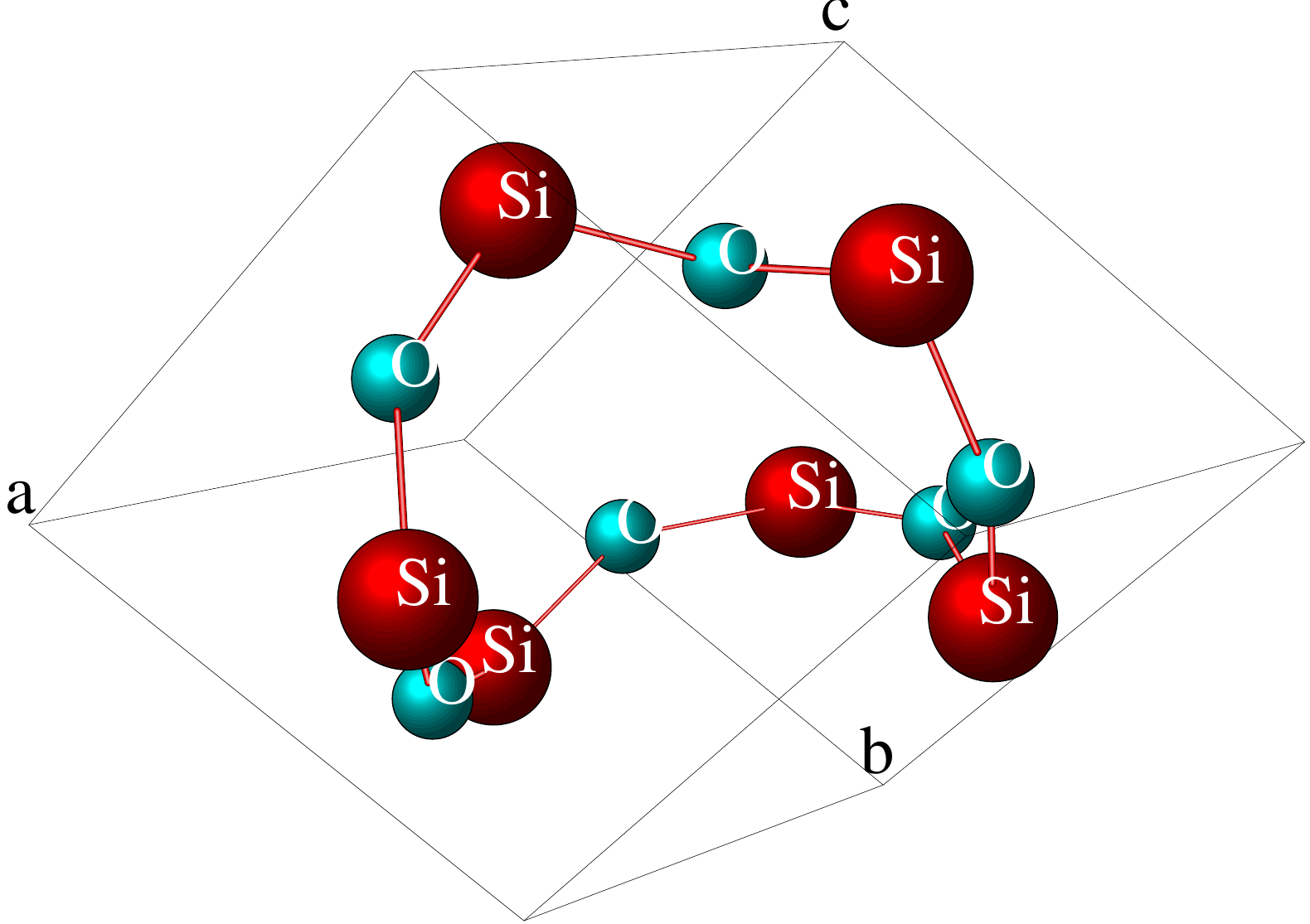}
\caption{Solid state unit cell of $\alpha$-quartz SiO$_2$. 
$a$=   4.9134 \AA, $b$= 4.9134 \AA, $c$= 5.4053 \AA, 
angles $({\bm a},{\bm b})$= 120\deg,  $({\bm a},{\bm c})$= 90 \deg, $({\bm b},{\bm c})$ = 90 \deg.
Figure drawn with Steffen Weber JSV software.}
\label{fig1}
\end{figure}

Quartz exists in two varieties: left-handed and right-handed that are
mirror images of each other as displayed in fig.2. Handedness or Chirality implies that the two
varieties have the same lattice energy (crystal energy of formation) and lack of center symmetry within each 
variety indicates that they belong to non-centro-symmetric groups as explained in the appendix.

\begin{figure}[htbp]
\includegraphics[width=2.5in]{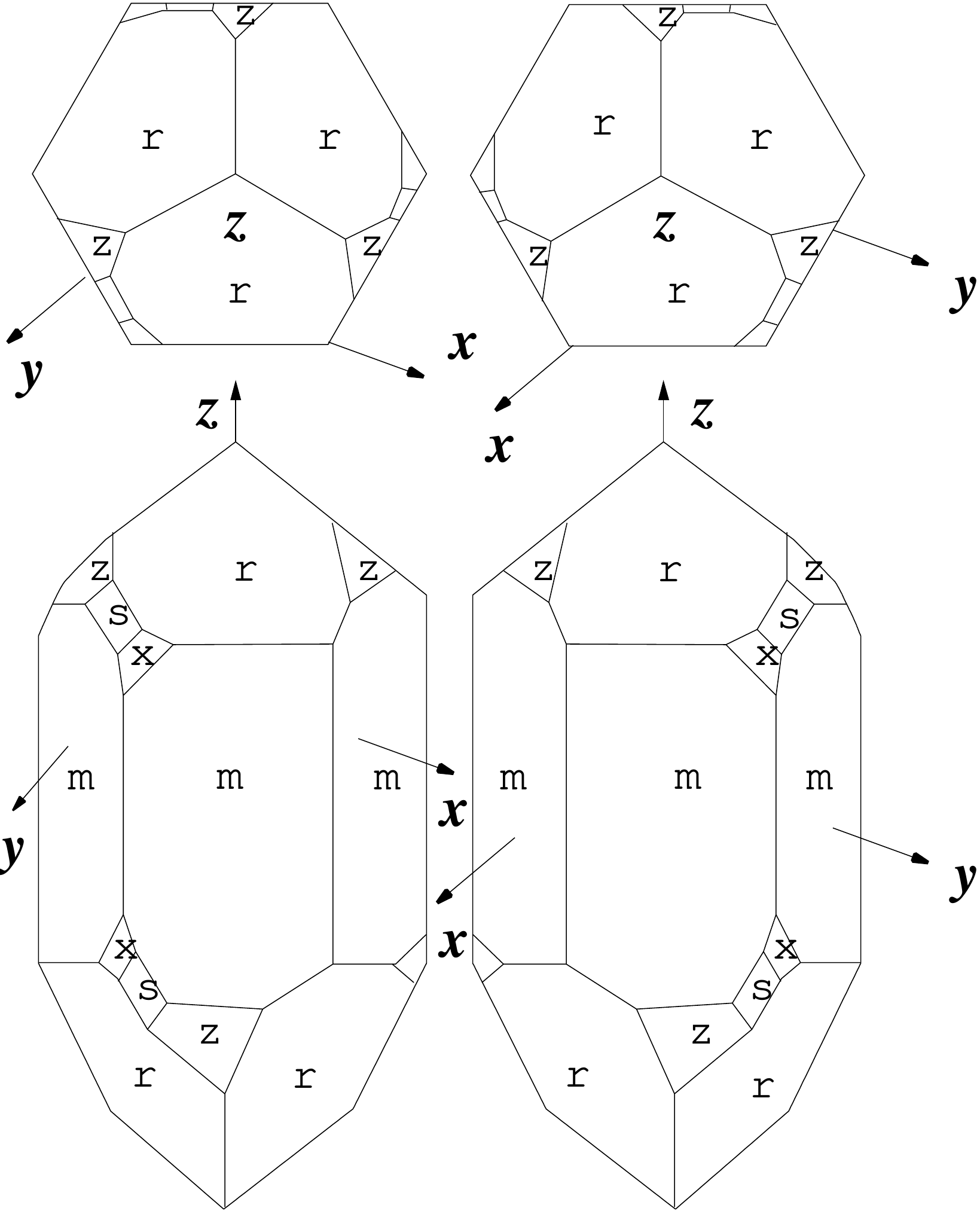}
\caption{Left-handed and right-handed types of Quartz with x,r,s,z and m faces.
In the left-hand case, an x- or s-face is present on the left side of a top r-face.
The right-hand type is when an x- or s-face is present on the right side of a top r-face.}
\label{fig2}
\end{figure}

The technological importance of Quartz originates from the values of its quality factor $Q$ 
that indicates the sharpness of resonance and 
electro-mechanical coupling coefficient $K$ that determines 
the conversion efficiency of mechanical into electrical energy and vice versa
as compared with other materials as seen in fig.3.

\begin{figure}[htbp]
\includegraphics[width=2.5in]{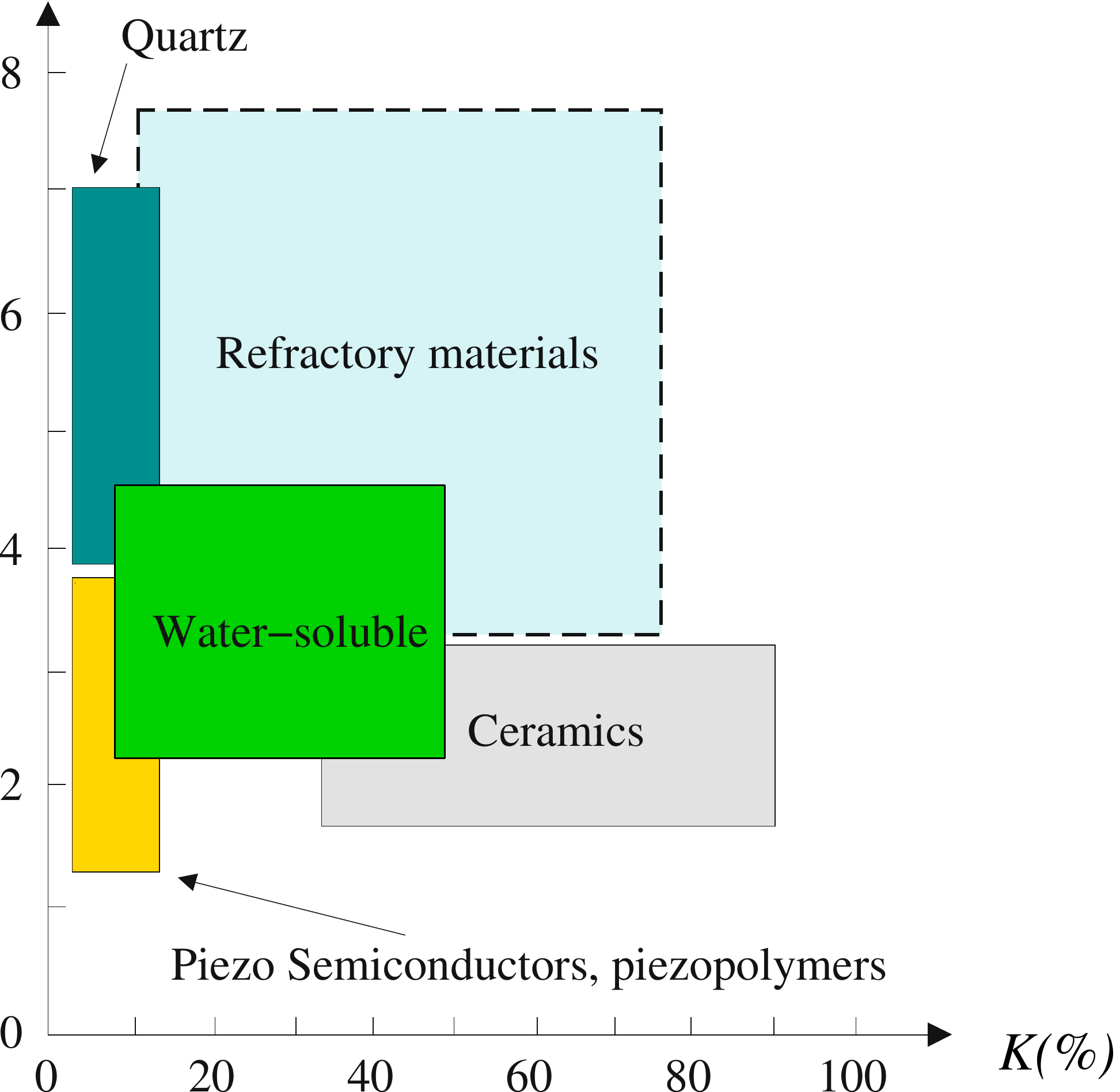}
\caption{Logarithm of the quality factor $\log_{10}Q$ and electro-mechanical coupling coefficient $K$ of Quartz and other materials indicating the importance of Quartz as originating from its high $Q$ value. Adapted from Ballato~\cite{Ballato}.}
\label{fig3}
\end{figure}

Piezoelectricity is a fundamental property of Quartz and is found in non-centrosymmetric
crystals that occur in two types of point symmetry groups (PSG) (see fig.~\ref{centro}).
There are ten PSG called polar groups (possessing a special direction) associated with 
pyroelectric and piezoelectric materials (possessing spontaneous polarization along the special direction) and
ten other  PSG that are piezoelectric only (their polarization being induced by mechanical deformation).

These PSG are classified as polar and non polar:

\begin{itemize}
\item Pyroelectric and piezoelectric (Polar groups displaying spontaneous polarization along a special
direction):

         \begin{itemize}
              \item Triclinic system ${\bm C}_1$
\item Monoclinic system ${\bm C}_s$, ${\bm C}_{2}$ 
\item Orthorhombic system  ${\bm C}_{2v}$
\item Tetragonal system ${\bm C}_{4}$, ${\bm C}_{4v}$
\item Trigonal (Rhombohedric) system ${\bm C}_{3}$, ${\bm C}_{3v}$
\item Hexagonal system  ${\bm C}_{6}$, ${\bm C}_{6v}$
           \end{itemize}

\item Piezoelectric only (Non polar groups characterized by a polarization induced by mechanical deformation):
         \begin{itemize}
              \item Orthorhombic system  ${\bm D}_2$
\item Tetragonal system ${\bm D}_4$, ${\bm D}_{2d}$, ${\bm S}_4$ 
\item Trigonal (Rhombohedric) system ${\bm D}_3$
\item Hexagonal system  ${\bm D}_6$, ${\bm C}_{3h}$, ${\bm D}_{3h}$
\item Cubic system  ${\bm T}$, ${\bm T}_d$
           \end{itemize}

\end{itemize}

Quartz belongs to  ${\bm D}_3$ group that possesses an order 3 rotation symmetry axis ($2\pi/3$ angle) that we might take along ${\bm z}$ axis. This axis has the  ${\cal R} ({\bm z},2\pi/3)$ rotation symmetry operation as well as  three order 2 axes  ($\pi$ rotation  symmetry) in the $xy$ plane.
The coordinate system we use is cartesian with basis
vectors $\bm{e}_1,\bm{e}_2, \bm{e}_3 $ such that $ \bm{e}_i \cdot \bm{e}_j=\delta_{ij}$ 
and any vector is expressed in this basis
as: $\bm{r}=x \bm{e}_1 + y \bm{e}_2 + z \bm{e}_3$.  \\
Neumann's principle states that 
"Symmetry elements of any physical property of a crystal must include the symmetry elements of the point group of the crystal" implying that crystal physical quantities are preserved after performing point group symmetry operations on them.

A symmetry operation such as a rotation by an angle $\phi$ about the ${\bm z}$ axis
denoted by  ${\cal R} ({\bm z},\phi)$ and represented by:
\be \begin{pmatrix} x' \cr y' \cr z' \end{pmatrix} = \begin{pmatrix} \cos{\phi} & -\sin{\phi} & 0 \cr \sin{\phi} & \cos{\phi} & 0 \cr
 0 & 0 & 1 \end{pmatrix} \begin{pmatrix} x \cr y \cr z \end{pmatrix}
\label{transform} 
\ee

transforms $\bm{r}=x \bm{e}_1 + y \bm{e}_2 + z \bm{e}_3$ into 
$\bm{r'}=x' \bm{e}_1 + y' \bm{e}_2 + z' \bm{e}_3$. This is different from
the case of rotation with basis change implying that the 
transformed vector $\bm{r'}=x' \bm{e'}_1 + y' \bm{e'}_2 + z' \bm{e'}_3$ is expressed in the rotated basis
$(\bm{e'}_1, \bm{e'}_2, \bm{e'}_3)$ such that:
\be \begin{pmatrix} x' \cr y' \cr z' \end{pmatrix} = \begin{pmatrix} \cos{\phi} & \sin{\phi} & 0 \cr -\sin{\phi} & \cos{\phi} & 0 \cr
 0 & 0 & 1 \end{pmatrix} \begin{pmatrix} x \cr y \cr z \end{pmatrix}
\label{basis_change} 
\ee

We examine below symmetry operations and implications of Neumann's principle in order to simplify 
the piezoelectric coefficients.

\section{Evaluation of piezoelectric coefficients by Fumi method}
Piezoelectric coefficients are represented by $d_{i,jk}$ a rank 3  tensor with 
indices $i,j,k=1,2,3$ corresponding to $x \rightarrow 1, y\rightarrow 2, z \rightarrow 3$. 
They originate from the relation ${\bm P}_i= d_{i,jk} \sigma_{jk}$ linking polarization vector 
${\bm P}$ to stress tensor $\sigma_{jk}$. \\

In total, we have 27 $d_{i,jk}$ coefficients since $i,j,k=1,2,3$, however 
writing $d_{i,jk}$ means index $i$ must be treated separately from indices
$j,k$ since $i$ relates to polarization ${\bm P}$ whereas $j,k$ indices relate to
the symmetric stress tensor $\sigma$ i.e. $\sigma_{jk}$=$\sigma_{kj}$.\\

The $j,k$ symmetry is exploited with Voigt notation that amounts to replace two indices 
by a single one according to the recipe: when $j=k, (j,k) \rightarrow j$ and
when $j\ne k, (j,k) \rightarrow 9-(j+k)$. More specifically, we have six possibilities:
$11 \rightarrow 1, 22 \rightarrow 2, 33 \rightarrow 3, 23 \rightarrow 4, 31 \rightarrow 5, 12 \rightarrow 6$. \\ 
The total number of $d_{i,jk}$ coefficients is 18 since $i=1,2,3$ and
Voigt index has six possibilities.

As a result, the Voigt piezoelectric matrix is $6\times 3$ with the explicit entries: 
\be 
\begin{pmatrix} d_{11} & d_{12} & d_{13} & d_{14} & d_{15} & d_{16}\cr
 d_{21} & d_{22} & d_{23} & d_{24} & d_{25} & d_{26}\cr
 d_{31} & d_{32} & d_{33} & d_{34} & d_{35} & d_{36} \end{pmatrix} 
\ee

where elements whose Voigt index is 4,5,6 are given by: $d_{i4}=d_{i23}+d_{i32}$, $d_{i5}=d_{i31}+d_{i13}$, 
$d_{i6}=d_{i12}+d_{i21}$ for $i=1,2,3$ as a result of symmetry.

Appendix B lists symmetry operations proper to each symmetry group. Quartz $D_3$ trigonal group
has the symmetry operations:  $E$,  2$C_3$,  3$C_2$.
The 3-fold rotation by $2\pi/3$ about ${\bm z}$ axis is denoted ${\cal R} ({\bm z},2\pi/3)$ and the 2-fold rotation by $\pi$ about the ${\bm x}$ axis is denoted ${\cal R} ({\bm x},\pi)$. \\

In order to perform symmetry transformations on the $d_{i,jk}$ coefficients, we apply the Italian physicist Fausto G. Fumi~\cite{Fumi} rule that states they transform as $x_i, x_j x_k$
written symbolically as $d_{i,jk} \sim x_i, x_j x_k$ with the condition of respecting the order of the
corresponding factors.\\

We start by considering rotational symmetry of order 2 about $x$ or ${\cal R} ({\bm x},\pi)$ operations: \\

The relationship between the rotated axes and the original axes in the 2-fold rotation about ${\bm x}$ is given as:
\be x'  = x, y'  = -y, z'  = -z \ee

Let us consider the implications of this mapping on some tensor elements.

$d_{111}$ transforms as:
\be x'x'x'  = x x x\ee

Thus $d_{111}'= d_{111}$  by Neumann's Principle, implying $d_{11} \ne 0$.

Coefficient $d_{211}$ transforms as:
\be  y'x'x'  = (-y) x x\ee

Thus $d_{211}'= -d_{211} = d_{211}$ by Neumann's Principle, implying $d_{211}=0$ or $d_{21} = 0$.

From this result, we infer that tensor elements with odd number of indices 2 and 3 are 0 by the 2-fold rotation (because of the transformation $y, z \rightarrow -y, -z $ ). \\ 
Hence, fourteen coefficients
$d_{113}, d_{131}$, 
$d_{112}, d_{121}$, 
$d_{211}, d_{222}$, 
$d_{233}, d_{223}$, 
$d_{232}, d_{311}$,
$d_{322}, d_{333}$, 
$d_{323}, d_{332}$ 
are zero along with their ten Voigt equivalents
$d_{15}, d_{16}, d_{21}, d_{22}, d_{23}, d_{24}, d_{31}, d_{32}, d_{33}, d_{34}$

As a result, the piezoelectric matrix is written as:
\be \begin{pmatrix} d_{11} & d_{12} & d_{13} & d_{14} & 0 & 0\cr 0 & 0 & 0 & 0 & d_{25} & d_{26}\cr
 0 & 0 & 0 & 0 & d_{35} & d_{36} \label{coefficients} \end{pmatrix} \ee

From the initial 18 coefficients only  8 coefficients  $d_{11}, d_{12}, d_{13}, d_{14}, d_{25}, d_{26}, d_{35}, d_{36}$ survive after the ${\cal R} ({\bm x},\pi)$ symmetry.\\

Let us investigate the impact of rotational symmetry of order 3 about ${\bm z}$ axis or 
${\cal R} ({\bm z},2\pi/3)$ on these coefficients.\\

For the 3-fold rotation about $z$, we use eq.~\ref{transform} with $\phi=\frac{2 \pi}{3}$ to express the relationship between the rotated coordinates and the original ones:
\be  x'  = -\frac{1}{2} x - \frac{\sqrt{3}}{2} y,  y'  = \frac{\sqrt{3}}{2} x - \frac{1}{2} y,  z'  =  z  \ee

\begin{itemize}

\item   Coefficient $d_{11}$, or $d_{111}$ transforms as:
\bea  x'x'x' & = & (-\frac{1}{2} x - \frac{\sqrt{3}}{2} y)(-\frac{1}{2} x - \frac{\sqrt{3}}{2} y)(-\frac{1}{2} x - \frac{\sqrt{3}}{2} y)\cr
 & = & -\frac{1}{8} x x x - \frac{\sqrt{3}}{8} x x y - \frac{\sqrt{3}}{8} x y x - \frac{3}{8} x y y \cr
 & & - \frac{\sqrt{3}}{8} y x x - \frac{3}{8} y x y - \frac{3}{8} y y x - \frac{3\sqrt{3}}{8} y y y\eea

This implies: $d_{111}' =-\frac{1}{8} d_{111} - \frac{3}{8} d_{122} - \frac{3}{8} d_{212} - \frac{3}{8} d_{221}$.

Moving on to Voigt notation and using Neumann's Principle, we have:
\be d_{11}' = -\frac{1}{8} d_{11} - \frac{3}{8} d_{12} - \frac{3}{8} d_{26} \equiv d_{11} \ee
resulting in: $3d_{11} + d_{12} + d_{26} = 0$.

\item Coefficient $d_{12}$ or $d_{122}$ transforms as:
\bea  x'y'y' & = &(-\frac{1}{2} x - \frac{\sqrt{3}}{2} y)(\frac{\sqrt{3}}{2} x - \frac{1}{2} y)(\frac{\sqrt{3}}{2} x - \frac{1}{2} y)\cr
 & = & -\frac{3}{8} x x x - \frac{\sqrt{3}}{8} x x y  + \frac{\sqrt{3}}{8} x y x - \frac{1}{8} x y y \cr
 & & - \frac{3\sqrt{3}}{8} y x x + \frac{3}{8} y x y  + \frac{3}{8} y y x - \frac{\sqrt{3}}{8} y y y\eea

yielding: $d_{122}' = -\frac{3}{8} d_{111} - \frac{1}{8} d_{122} + \frac{3}{8} d_{212} + \frac{3}{8} d_{221}$.

Using Voigt notation and Neumann's Principle, we get:
\be d_{12}' = -\frac{3}{8} d_{11} - \frac{1}{8} d_{12} + \frac{3}{8} d_{26} \equiv d_{12}\ee 
implying:  $d_{11} + 3 d_{12} - d_{26} = 0$. \\

Combining relations $3d_{11} + d_{12} + d_{26} = 0$ and $3d_{11} + d_{12} + d_{26} = 0$, we get:
$d_{12} = -d_{11}$  and $d_{26} = -2d_{11}$.

\item Coefficient $d_{13}$ or $d_{133}$ transforms as:

\bea  x'z'z' & = &(-\frac{1}{2} x - \frac{\sqrt{3}}{2} y)zz\cr     
             & =&  -\frac{1}{2} x z z - \frac{\sqrt{3}}{2} y z z \eea

This yields $d_{133}'  = -\frac{1}{2} d_{133} - \frac{\sqrt{3}}{2} d_{233}$ and consequently
 $ d_{133}'  = -\frac{1}{2} d_{133}$ since $d_{233}=0$.
Using Neumann's Principle,  $d_{133}'=-\frac{1}{2} d_{133} \equiv d_{133}$ implies $d_{133}  = 0$ and 
consequently $d_{13}  = 0$.

\item  Coefficients $d_{14}$ and $d_{25}$ : \\
$d_{213}$ being part of $d_{25}=d_{213}+d_{231}$  transforms as:
\bea  y'x'z' & = &(\frac{\sqrt{3}}{2} x - \frac{1}{2} y)(-\frac{1}{2} x - \frac{\sqrt{3}}{2} y)z\cr 
& = & -\frac{\sqrt{3}}{4} x x z - \frac{3}{4} x y z + \frac{1}{4} y x z + \frac{\sqrt{3}}{4} y y z \eea

Thus $d_{213}'  = -\frac{\sqrt{3}}{4} d_{113} - \frac{3}{4} d_{123} + \frac{1}{4} d_{213} 
+ \frac{\sqrt{3}}{4} d_{223} $, 
implying $d_{213}'  = -\frac{3}{4} d_{123} + \frac{1}{4} d_{213} \equiv d_{213} $. \\

This leads to $ d_{213}= -d_{123} $ implying $d_{25}  = - d_{14}$.  \\

\item  Coefficient $d_{313}$ being a part of $d_{35}$ i.e. $d_{313}+d_{331}$ transforms as:

\bea
  z'x'z' & = & z(-\frac{1}{2} x - \frac{\sqrt{3}}{2} y)z \nonumber \\
 & = & -\frac{1}{2} z x z - \frac{\sqrt{3}}{2} z y z \nonumber \\
\eea 
This gives $ d_{313}'  = -\frac{1}{2} d_{313} - \frac{\sqrt{3}}{2} d_{323}$ 
resulting in $d_{313}'  = -\frac{1}{2} d_{313}$ since $d_{323}$ =0.
Therefore we get with Neumann's Principle $d_{313}  = 0$ and  $d_{331} = 0$ implying $d_{35}= 0$. \\

\item Coefficient $d_{312}$ being a part of $d_{36}$ or $d_{312}+d_{321}$ transforms as:
\bea z'x'y' & = & z(-\frac{1}{2} x - \frac{\sqrt{3}}{2} y)(\frac{\sqrt{3}}{2} x - \frac{1}{2} y)\cr 
& = & -\frac{\sqrt{3}}{4} z x x + \frac{1}{4} z x y - \frac{3}{4} z y x + \frac{\sqrt{3}}{4} z y y \eea

thus $d_{312}'  = -\frac{\sqrt{3}}{4} d_{311} + \frac{1}{4} d_{312} - \frac{3}{4} d_{321} + \frac{\sqrt{3}}{4} d_{322}$

This yields:

\be d_{312}'  = \frac{1}{4} d_{312} - \frac{3}{4} d_{321} \equiv d_{312}\ee
that is:  $d_{312}  = - d_{321}$ implying $d_{36} = d_{312} + d_{321} = 0$. \\

\end{itemize}

Collecting all coefficients the piezoelectric matrix becomes:

\be \begin{pmatrix} d_{11} & -d_{11} & 0 & d_{14} & 0 & 0\cr
 0 & 0 & 0 & 0 & -d_{14} & -2d_{11}\cr
 0 & 0 & 0 & 0 & 0 & 0  \label{final} \end{pmatrix} \ee

\section{Evaluation of piezoelectric coefficients by Landau-Lifshitz method}

Instead of working with matrices while applying Fumi~\cite{Fumi} recipe to the transformation  
of coefficients $d_{i,jk} \sim x_i; x_j x_k$, we recall that performing rotation operations in a plane
orthogonal to $z$ may be described by complex variables as done in Landau-Lifshitz book~\cite{Landau}: \\
 $\xi \rightarrow \xi e^{\frac{2i\pi}{3}}$
$\eta \rightarrow \eta  e^{- \frac{2i\pi}{3}}$,  $z \rightarrow z$. \\

We can use new complex variables $\xi,\eta$ in the $xy$ plane through the variable change: $\xi=x+iy$, $\eta=\xi^{*}=x-iy$.
Note that the variable sets $\xi,\eta$ as well as $xy$ are linearly independent (possessing a non-zero
Wronskian)~\cite{Landau}.

\begin{enumerate}
\item Rotational symmetry of order 3 about ${\bm z}$ axis or ${\cal R} ({\bm z},2\pi/3)$ operations: \\ 
Index separation  $i,jk$ yields by Fumi~\cite{Fumi} rule, 
terms such as $({z, \xi \eta}), ({\eta, z \xi}),({\xi,  z \eta})$. The transformation
applies in the same manner to complex phases: \\
$x_i \rightarrow x_i e^{i\phi_i}, x_j \rightarrow x_j e^{i\phi_j}, x_k \rightarrow x_k e^{i\phi_k}$  
obtaining $(x_i,x_j x_k) \rightarrow (x_i,x_j x_k) e^{i(\phi_i +\phi_j + \phi_k)}$ where
 $(x_i,x_j x_k)$ coordinates represent $({z, \xi \eta})$. \\

The transformation $d_{z,z\xi}  \rightarrow  d_{z,z\xi} e^{ \frac{2i\pi}{3}}$ along with
${\bm D}_3$ symmetry (invariance with respect to rotation ${\cal R} ({\bm z},2\pi/3)$
implies: $d_{z,z\xi} =  d_{z,z\xi} e^{ \frac{2i\pi}{3}}$ which results in: 
$d_{z,z\xi} (1- e^{ \frac{2i\pi}{3}})= 0$, thus $d_{z,z\xi}=0$. 
Similarly $d_{\xi,zz}$, $d_{\eta,zz}$ are zero since total phase 
would be $\pm \frac{2\pi}{3}$,
same for $d_{z,\xi \xi}$, $d_{z;\eta \eta}$ for which the phase is $\pm \frac{4\pi}{3}$. \\

The non-zero terms invariant with respect to
${\cal R} ({\bm z},2\pi/3)$ should contain combination of 
$({z, \xi \eta}), ({\eta, z \xi}),({\xi,  z \eta}),({\xi, \xi \xi}),({\eta, \eta \eta}),({z,zz})$
since the total phase obtained after  ${\cal R} ({\bm z},2\pi/3)$ operation is 0 or $\pm 2\pi$.

Finally the 6 non-zero terms correspond to the combination: 
$d_{z, \xi \eta}, d_{\eta, z \xi},d_{\xi,  z \eta},d_{\xi, \xi \xi},d_{\eta, \eta \eta},d_{z,zz}$. Note the existence of $d_{z, \xi \eta}, d_{\eta, z \xi},d_{\xi,  z \eta}$ 
terms with $z$ appearing only once~\cite{Fumi}.

\item Rotational symmetry of order 2 about $x$ or ${\cal R} ({\bm x},\pi)$ operations: \\

This symmetry is carried out through the following transformations: \\
$x \rightarrow x, y \rightarrow -y, z \rightarrow -z$ i.e.
$\xi \rightarrow \eta, \eta \rightarrow \xi, z \rightarrow -z$ . \\ 
This eliminates all terms containing  an odd number of $z$ such as $d_{z, \xi \eta},d_{z,zz}$. \\

\end{enumerate}

Applying transformation to $d_{\eta, z \xi}$, we get  $d_{\xi,-z \eta}$
or  $-d_{\xi,z \eta}$ thus term $d_{\eta, z \xi}$ is not zero (resulting from changing
index $\eta$ into $\xi$), whereas
$d_{z, \xi \eta}$ transforms into $d_{-z, \eta \xi}$ or $-d_{z, \xi \eta}$ thus this term
is zero.\\ 

Finally only 2 terms  $d_{\eta, z \xi}$ and $d_{\xi, \xi \xi}$ remain 
since we have:
$d_{\eta, z \xi}= -d_{\xi, z \eta}$ and $d_{\xi, \xi \xi}=d_{\eta, \eta \eta}$. \\

Going back to $x,y$ variables from $\xi,\eta$, we use energy conservation
in order to avoid problems stemming from non-orthogonality of coordinate system $(z, \xi,\eta)$ 
in contrast to  $(z,x,y)$ orthogonal system.\\

The energy of the system is given by  $(z, \xi,\eta)$ par $-{\bm {E.P}}= -E_i P_i= -E_i d_{i,jk} \sigma_{jk}$, obtaining: \\ 
$-{\bm {E.P}} = -2 d_{\eta, z \xi} (E_{\eta} \sigma_{z \xi} - E_{\xi} \sigma_{z \eta})
- d_{\xi, \xi \xi} (E_{\xi} \sigma_{\xi \xi} + E_{\eta} \sigma_{\eta \eta})$. \\

We apply Fumi~\cite{Fumi} rule to transform energy in  
system $(z,x,y)$ using correspondence between indices and tensor components as 
follows: $E_{\xi}= E_x+i E_y$, $E_{\eta}= E_x -i E_y$.
Similarly,   $\xi\xi=xx-yy+2ixy$ should yield stress
tensor $\sigma$ components as $\sigma_{\xi\xi}=\sigma_{xx}-\sigma_{yy}+2i \sigma_{xy}$ whereas 
$\xi\eta=xx+yy$ should yield: $\sigma_{\xi\eta}=\sigma_{xx}+\sigma_{yy}$ and so forth. \\

The energy writes: 
$2 a (E_{y} \sigma_{z x} - E_{x} \sigma_{z y}) +b [2 E_{y} \sigma_{xy} - E_{x}(\sigma_{xx}-\sigma_{yy})]$,
with real constants $a,b$ defined by $a= 2i d_{\eta, z \xi}$, and $b=2 d_{\xi, \xi \xi}$. 
As a result, we have the remaining components
$d_{x,yz}= -d_{y,zx}= a$ and $d_{y,xy}= -d_{x,xx}=d_{x,yy}=-b$. \\

Collecting all terms, the matrix becomes:

$$
\begin{pmatrix} b & -b & 0 & a & 0 & 0 \cr
         0 & 0& 0& 0 & -a & -b  \cr   
         0& 0& 0 & 0 & 0 & 0   
\end{pmatrix}
$$

Moving on to Voigt representation, we transform: \\
$d_{i,jk} \rightarrow d_{i,jj}, j=1,2,3 $ when $j=k$ whereas 
$d_{i,jk} \rightarrow 2d_{i,jk} $ for terms $j \ne k$, since we have to account for coefficients 
symmetry: $yz \leftrightarrow zy,xz \leftrightarrow zx,xy \leftrightarrow yx$.

Thus we obtain:

$$
\begin{pmatrix}b & -b & 0 & 2a & 0 & 0 \cr
         0 & 0& 0& 0 & -2a & -2b  \cr   
         0& 0& 0 & 0 & 0 & 0   
\end{pmatrix}
$$

that might be written in the form shown in eq.~\ref{final} which is exactly the result obtained 
previously by Fumi method.

\section{Evaluation of piezoelectric coefficients by Royer-Dieulesaint method}
Royer-Dieulesaint~\cite{Royer} method is the most elegant. It is based on dealing with rotation matrices 
through their eigenvalues which classifies this method as an intermediate between Fumi and Landau-Lifshitz.

After performing 2-fold rotation about $x$ axis, we infer as before that tensor elements with odd number of indices 2 and 3 are 0 from the transformation $y, z \rightarrow -y, -z $ ). \\ 

As a result, the piezoelectric matrix is written as in eq.~\ref{coefficients}.\\

In order to tackle the ${\cal R} ({\bm z},\phi)$ transformation, we start with the corresponding general rotation matrix given in eq.~\ref{transform} with $\phi=\frac{2\pi}{n}$ with $n$ an integer. \\

The eigenvalues of this matrix are: 
$\lambda_1=e^{i\phi},\lambda_2=e^{-i\phi}, \lambda_3=1$ and the corresponding eigenvectors are given by:
$\bm{\xi}_1=\left(\frac{1}{\sqrt{2}},\frac{i}{\sqrt{2}},0 \right), 
\bm{\xi}_2=\left(\frac{i}{\sqrt{2}},\frac{1}{\sqrt{2}},0 \right),
\bm{\xi}_3=(0,0,1)$.

From the eigenvectors we derive the transformation matrix that takes us from the
$(\bm{\xi}_1, \bm{\xi}_2, \bm{\xi}_3)$ to the initial orthonormal basis 
$\bm{e}_1,\bm{e}_2, \bm{e}_3 $ such as:

\be
A= \begin{pmatrix} \frac{1}{\sqrt{2}} &\frac{i}{\sqrt{2}} & 0  \cr
         \frac{i}{\sqrt{2}}&\frac{1}{\sqrt{2}} &0 \cr   
         0 & 0 & 1   \end{pmatrix} 
\label{Amatrix}
\ee

In the $(\bm{\xi}_1, \bm{\xi}_2, \bm{\xi}_3)$ basis, the piezoelectric coefficient tensor
is written as $\eta_{ijk}$ and the transformation from $d_{ijk}$ to $\eta_{ijk}$ is given by 
$d_{ijk}=A_{il}A_{jm}A_{kn} \eta_{lmn}$.

A symmetry transformation combined with Neumann's Principle yields:

\be \eta_{ijk}= \lambda_i \lambda_j \lambda_k \eta_{ijk} \ee

If we call $\nu_1$ the number of indices equal to 1 and  $\nu_2$ 
the number of indices equal to 2, $\lambda_i \lambda_j \lambda_k=\exp[i(\nu_1-\nu_2)\frac{2\pi}{n}]$ for a symmetry axis of order $n$. In the ${\cal R} ({\bm z},2\pi/3)$ symmetry $n=3$ and $\eta_{ijk}$ components  are not zero whenever 
$\nu_1-\nu_2$ is a multiple of $n$. This implies that 
$\eta_{123}$, $\eta_{213}$, $\eta_{312}$ and $\eta_{333}$ are not zero since
$\nu_1-\nu_2=0$ as well as components $\eta_{111}$ and $\eta_{222}$ since $\nu_1-\nu_2= \pm 3$.

Let us consider first $d_{ijk}$ case with $i,j,k \ne 3$  such that coefficients are expressed in terms of $\eta_{111}$ and $\eta_{222}$ only.

Thus $d_{ijk}=A_{i1}A_{j1}A_{k1} \eta_{111}+ A_{i2}A_{j2}A_{k2} \eta_{222}$ for $i,j,k \ne 3$.

For instance, if we want to evaluate $d_{11}$ i.e. $d_{111}$, we write:
$d_{111}=A_{11}A_{11}A_{11} \eta_{111}+ A_{12}A_{12}A_{12} \eta_{222}=0$.
Using matrix $A$ elements given in eq.~\ref{Amatrix} we get:
$d_{111}=\frac{1}{2\sqrt{2}} \eta_{111}-\frac{i}{2\sqrt{2}} \eta_{222}$.

Moving on to $d_{12}$ i.e. $d_{122}$, we obtain in the same way:
$d_{122}=-\frac{1}{2\sqrt{2}} \eta_{111}+\frac{i}{2\sqrt{2}} \eta_{222}$ which
implies that $d_{11}=-d_{12}$.

In the same manner we can evaluate $d_{26}$ i.e. $d_{212}$ or $d_{221}$.
We obtain $d_{212}=-\frac{1}{2\sqrt{2}} \eta_{111}+\frac{i}{2\sqrt{2}} \eta_{222}$
which implies that $d_{26}=-2d_{11}$ (factor 2 originates from the fact
$d_{26}$ is equivalent to $d_{212}$ or $d_{221}$ as previously done 
in the Landau-Lifshitz section).

Elements containing digit 3 are $d_{13}$, $d_{35}$ and  $d_{36}$. 
They are all zero as we know from Fumi analysis. 
Let us retrieve this result in the case of $d_{13}$ or $d_{133}$. \\

In order to evaluate $d_{133}=A_{1l}A_{3m}A_{3n} \eta_{lmn}$, we  use $A_{13}=A_{23}=A_{31}=A_{32}=0$, obtaining: $d_{133}=A_{11}A_{33}A_{33} \eta_{133}+A_{12}A_{33}A_{33} \eta_{233}=0 $
since both $\eta_{133}$ and $\eta_{233}$ are zero.

In the $d_{35}$ case, one has to evaluate $d_{313}$ and $d_{331}$. Evaluating 
$d_{313}= A_{3l}A_{1m}A_{3n} \eta_{lmn}$ yields 
$d_{313}= A_{33}A_{11}A_{33} \eta_{313} + A_{33}A_{12}A_{33} \eta_{323}$  which is zero
since both $\eta_{313}$and $\eta_{323}$ are zero.

The rest of the elements are obtained in the same fashion and the final outcome is exactly what we obtained earlier from Fumi and Landau-Lifshitz albeit in a faster and more compact form. 

The piezoelectric coefficient matrix obtained is the same as Fumi and Landau-Lifshitz 
previous result given in eq.~\ref{final}.

For right-handed $\alpha-$quartz, the actual numerical\cite{Cady} values are (each should be multiplied by $10^{-12}$  in order to get (Coulomb/Newton) SI units):

\be
\begin{pmatrix} -2.3 & 2.3 & 0 & -0.67 & 0 & 0 \cr
         0 & 0& 0& 0 & 0.67 & 4.6  \cr   
         0& 0& 0 & 0 & 0 & 0   \end{pmatrix}  
\ee

\section{Conclusion}
Symmetry is illustrated in the the evaluation of piezoelectric coefficients of
$\alpha$-Quartz by three distinct methods: Fumi, Landau-Lifshitz ans Royer-Dieulesaint.
Fumi method is general, straightforward and tedious, Landau-Lifshitz method requires
many physical concepts that must be adapted to every encountered situation whereas
Royer-Dieulesaint is the most elegant while general and not requiring any additional 
concepts as with Landau-Lifshitz. Advanced methods to deal with symmetry are based on
Group theoretical description of Tensors and Tensor fields such as described in ref.~\cite{Batanouny}, however they require deep knowledge of Group Theory~\cite{Dresselhaus}.

\newpage

\appendix

\section{Point symmetry groups and symmetry operations}

\begin{figure}[htbp]
\includegraphics[width=15cm]{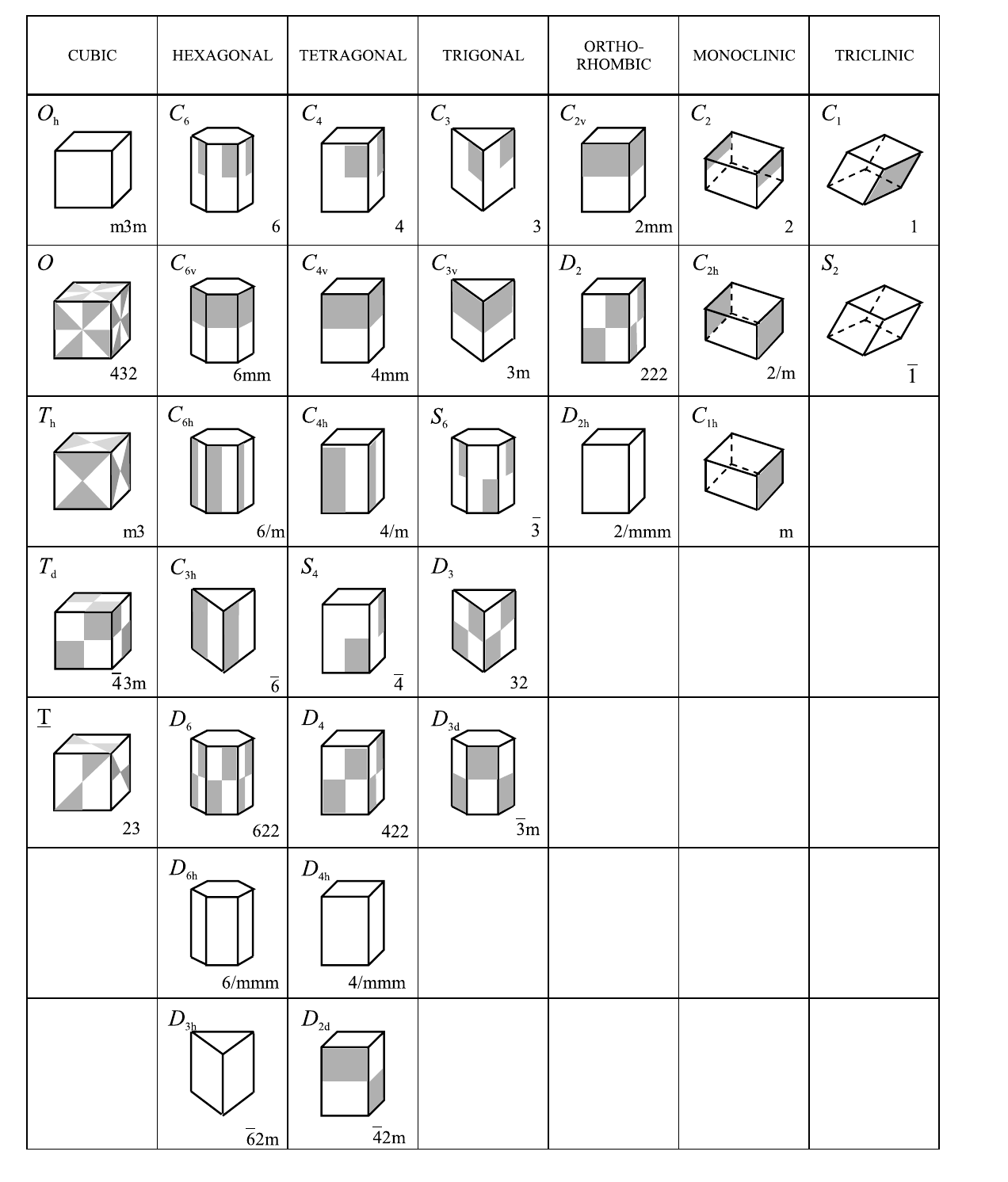}
\caption{Crystallographic point symmetry groups in 3D. Adapted from Ashcroft and Mermin~\cite{Ashcroft}.}
\label{PSG}
\end{figure}

\begin{figure}[htbp]
\includegraphics[width=15cm]{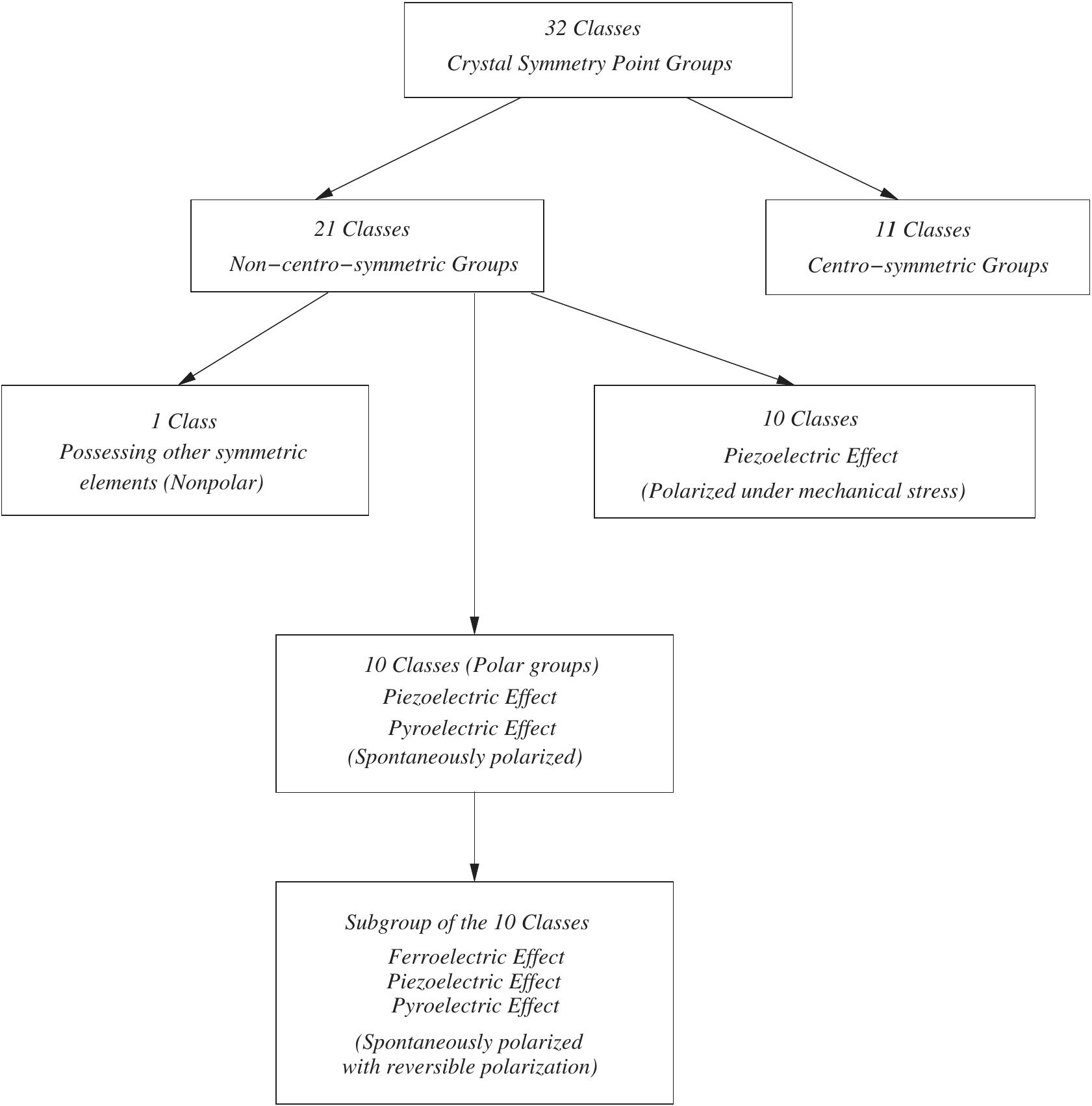}
\caption{Point symmetry groups  in which pyroelectric and 
piezoelectric effects are expected by lack of center symmetry. 
We have 10 groups in each case. Polar groups allow spontaneous 
polarization whereas in non-polar groups  polarization appears after 
application of stress. Adapted from Kao~\cite{Kao}.}
\label{centro}
\end{figure}

We first present 3D Point Symmetry Groups in fig.~\ref{PSG}. The classification into
centro, non centrosymmetric groups as well as polar and  non-polar groups is given in fig.~\ref{centro}.
Finally symmetry operations pertaining to each group is presented in Table I.

\begin{table}[htbp]
\begin{center}
\begin{tabular}{|K{20pt}|K{40pt}|K{120pt}|K{20pt}|}
\hline
(S) &   (H-M) & Symmetry operation & $N$ \\
\hline
\multicolumn{4}{c}{Cubic systems}   \\
\hline
  $T$  &   23 & $E$, 4$C_3$,  4$C_3^2$, 3$C_2$    & 12   \\
 $T_h$ & m3  & $E$,  8$C_3$,  3$C_2$,  3$\sigma_v$, $i$,  8$S_6$ &  24\\
 $O$ & 432  & $E$, 6$C_4$,  8$C_3$, 3$C_2$, 6$C^{'}_2$ & 24\\
 $T_d$ &   43m & $E$,  8$C_3$,  3$C_2$, 6$S_4$,  6$\sigma_d$ & 24\\
  $O_h$ & m3m  & $E$,  8$C_3$, 6$C_2$, 6$C_4$, 3$C^{'}_2$, $i$,  & \\
 &&  6$S_4$,  8$S_6$,  3$\sigma_h$, 6$\sigma_d$  & 48\\
\hline
\multicolumn{4}{c}{Tetragonal systems}\\
\hline
 $C_4$ & 4  & $E$,  $C_4$,  $C_2$,  $C_4^3$ & 4 \\
 $S_4$ & 4 & $E$,  $S_4$,  $C_2$,  $S_4^3$ & 4\\
$C_{4h}$ & 4/m & $E$,  $C_4$,  $C_2$,  $C_4^3$, $i$, $S_4^3$,  $\sigma_h$,  $S_4$ & 8\\
 $D_4$ & 422 & $E$,  2$C_4$,  2$C_2$,  2$C^{'}_2$, 2$C^{''}_2$ & 8 \\
$C_{4v}$ & 4mm & $E$,  2$S_4$,  $C_2$,  2$C^{'}_2$, 2$\sigma_d$  & 8 \\
 $D_{2d}$  & 42m & $E$,  2$S_4$,  $C_2$,  2$C^{'}_2$, 2$\sigma_d$ & 8\\
  $D_{4h}$ & 4/mmm & $E$,  2$C_4$, $C_2$, 2$C^{'}_2$, 2$C^{''}_2$, $i$,  & \\
 &&  2$S_4$,  $\sigma_h$,  2$\sigma_v$,  2$\sigma_d$ &  16 \\
\hline
\multicolumn{4}{c}{Orthorhombic systems} \\
\hline
$D_2$ & 222 & $E$,  $C_2$,  $C^{'}_2$,  $C^{''}_2$ & 4\\
 $C_{2v}$ & mm2  & $E$,  $C_2$,  $\sigma_v$,  $\sigma'_v$ & 4 \\
$D_{2h}$ & mmm & $E$,  $C_2$,  $C^{'}_2$,  $C^{''}_2$,  $i$,  $\sigma$,  $\sigma'$,  $\sigma''$ & 8 \\
\hline
\multicolumn{4}{c}{Monoclinic systems} \\
\hline
 $C_2$ & 2 & $E$,  $C_2$ & 2 \\
 $C_3$ & m or 2 & $E$,  $\sigma_h$  & 2  \\
 $C_{2h}$ &  2/m & $E$,  $C_2$,  $i$,  $\sigma_h$ & 4 \\
\hline
\multicolumn{4}{c}{Triclinic systems} \\
\hline
 $C_1$& 1 & $E$  & 1 \\
 $C_i$& 1  & $E$, $i$ & 2 \\
\hline
\multicolumn{4}{c}{Trigonal systems}\\
\hline
$C_3$ & 3 & $E$,  $C_3$,  $C_3^2$ & 3\\
 $S_6$ & 3 & $E$,  $C_3$,  $C_3^2$,  $i$,  $S_6^5$,  $S_6$  & 6 \\
 $D_3$ & 32 & $E$,  2$C_3$,  3$C_2$ & 6\\
$C_{3v}$ & 3m & $E$,  2$C_3$,  3$\sigma_v$ & 6 \\
 $D_{3d}$ &   3m  & $E$,  2$C_3$,  3$C_2$,  $i$,  2$S_5$,  3$\sigma_d$ & 12 \\
\hline
\multicolumn{4}{c}{Hexagonal systems}  \\
\hline
 $C_6$ & 6 & $E$,  $C_6$,  $C_3$,  $C_2$,  $C_3^2$,  $C_6^5$ & 6 \\
$C_{3h}$ ($S_3$) & 6 or 3/m & $E$,  $C_3$,  $C_3^2$,  $\sigma_h$,  $S_3$,  $S_3^5$ & 6\\
   $C_{6h}$ &   6/m  &  $E$,  $C_6$,  $C_3$,  $C_2$,  $C_3^2$,  $C_6^5$,  &\\
 &&  $i$, $S_3^5$, $S_6^5$,   $\sigma_h$,  $S_6$,  $S_3$  & 12\\
 $D_6$ & 622 & $E$,  2$C_6$,  2$C_3$,  $C_2$,  3$C^{'}_2$,  3$C^{''}_2$  & 12 \\
 $C_{6v}$ &  6mm & $E$,  2$C_6$,  2$C_3$,  $C_2$,   3$\sigma_v$,  3$\sigma_d$ & 12\\
 $D_{3h}$ & 6m2  & $E$,  2$C_3$,  3$C_2$,  $\sigma_h$,  2$S_3$,   3$\sigma_v$    & 12\\
 $D_{6h}$ & 6/mmm  & $E$,  2$C_6$,  2$C_5$,  $C_2$,  3$C^{'}_2$,  3$C^{''}_2$,  &\\
  &&  $i$,  2$S_3$,  2$S_6$,  $\sigma_h$,  3$\sigma_d$ , 3$\sigma_v$ & 24\\
\hline
\end{tabular}
\caption{Point symmetry groups in the Schoenflies (S) and  Hermann-Mauguin (H-M) nomenclature with corresponding symmetry operations and order $N$. The primed and double primed operations such as $C^{'}_3$ and $C^{''}_3$ correspond to 3-fold rotation by $2\pi/3$
with respect to axes other than the standard axis (usually $z$). $\sigma_v$ is a reflection operation with respect to a vertical plane (containing the $z$ axis) 
whereas $\sigma_h$ is reflection with respect to a horizontal plane and $\sigma_d$
is reflection with respect to a diagonal plane. The primed and double primed operations such as  $\sigma'$ and $\sigma''$ correspond to reflection operations with respect to
planes other than the standard plane.}
\end{center}
\label{operations}
\end{table}


\begin{thebibliography}{99}
\bibitem{Gross} D. Gross, Proc. Natl. Acad. Sci. USA {\bf 93}, 14256 (1996).
\bibitem{Landau} L. D. Landau and E. M. Lifshitz,  {\it Electrodynamics of Continuous Media},  Pergamon,  Oxford (1975).
\bibitem{Fumi} Fumi rule is based on the fact, components of an arbitrary tensor $A$ 
transform as product of the corresponding indices, i.e. $A_{ijkl...} \sim x_i x_j x_k x_l...$ 
Consequently $d_{i,jk}  \sim x_i, x_j x_k$; see F.G. Fumi Nuovo Cimento Vol. IX, 739 (1952).
\bibitem{Ballato} A. Ballato, chapter 2 in {\it Piezoelectricity: Evolution and future of a technology},
edited by W. Heywang, K. Lubitz and W. Wersing (Springer, New-York) (2008). 
\bibitem{Ashcroft} N. Ashcroft and D. Mermin {\it Solid State Physics}, (Holt, Rinehart and Winston, London) (1976). 
\bibitem{Kao} K. C. Kao {\it Dielectric phenomena in solids}, Elsevier, San Diego (2004).
\bibitem{Royer} D. Royer, E. Dieulesaint {\it  Elastic Waves in Solids I: Free and Guided Propagation}
Springer Science \& Business Media (1999).
\bibitem{Cady} W. G. Cady {\it Piezoelectricity: An introduction to the theory and applications
of electromechanical phenomena in crystals}, second edition, Dover, New-York (1964). see also
J. F. Nye, {\it  Physical properties of crystals and their representation by
tensors and matrices}, Oxford, new-York (1985).
\bibitem{Batanouny} M. El-Batanouny and F. Wooten,  {\it  Symmetry and Condensed Matter Physics, A Computational Approach}, Cambridge University Press, New-York (2008).
\bibitem{Dresselhaus} M.S. Dresselhaus, G. Dresselhaus and A. Jorio, {\it Group Theory
Application to the Physics of Condensed Matter}, Springer-Verlag, New-York (2008).
\end{thebibliography}
\end{document}